\documentclass[aps,showpacs,superscriptaddress,preprint]{revtex4}%
\usepackage{amsfonts}
\usepackage{amsmath}
\usepackage{amssymb}
\usepackage{graphicx}%
\begin{document}
\title{Electronic structure, mechanical and thermodynamic properties of ThN from
first-principles calculations}
\author{Yong Lu}
\affiliation{LCP, Institute of Applied Physics and Computational
Mathematics, Beijing 100088, People's Republic of China}
\affiliation{Department of Physics, Beijing Normal University,
100875, People's Republic of China}
\author{Da-Fang Li}
\affiliation{LCP, Institute of Applied Physics and Computational Mathematics, Beijing
100088, People's Republic of China}
\author{Bao-Tian Wang}
\affiliation{LCP, Institute of Applied Physics and Computational
Mathematics, Beijing 100088, People's Republic of China}
\affiliation{Institute of Theoretical Physics and Department of
Physics, Shanxi University, Taiyuan 030006, People's Republic of
China}
\author{Rong-Wu Li}
\affiliation{Department of Physics, Beijing Normal University, 100875, People's Republic of China}
\author{Ping Zhang}
\thanks{Author to whom correspondence should be addressed. E-mail: zhang\_ping@iapcm.ac.cn}
\affiliation{LCP, Institute of Applied Physics and Computational
Mathematics, Beijing 100088, People's Republic of China}
\affiliation{Center for Applied Physics and Technology, Peking
University, Beijing 100871, People's Republic of China}

\begin{abstract}
Lattice parameter, electronic structure, mechanical and
thermodynamic properties of ThN are systematically studied using the
projector-augmented-wave method and the generalized gradient
approximation based on the density functional theory. The calculated
electronic structure indicates the important contributions of Th
6\emph{d }and 5\emph{f} states to the Fermi-level electron
occupation. Through Bader analysis it is found that the effective
valencies in ThN can be represented as Th$^{+1.82}$ N$^{-1.82}$.
Elastic constant calculations shows that ThN is mechanically stable
and elastically anisotropic. Furthermore, the melting curve of ThN
is presented up to 120 GPa. Based on the phonon dispersion data, our
calculated specific heat capacities including both lattice and
conduction-electron contributions agree well with experimental
results in a wide range of temperature.

\end{abstract}
\maketitle

\section{introduction}

Actinide nitrides have been extensively studied in experiments in
connection with their potential applications in the Generation-IV
reactors \cite{Proc}. These reactors raise a number of concerns
surrounding the issue of nuclear power. The effective utilization of
nuclear power will require continued improvements in nuclear
technology, particularly related to safety and efficiency. Nowadays,
except for the oxide based fuels, the nitride fuels also participate
in the competition to become the alternative materials for their
superior thermophysical properties, such as high melting point, high
thermal conductivity, and high density, as well as the good
compatibility with the coolant (liquid sodium)
\cite{Matzke1992,Kleykamp,Evensen,Bauer}. Since the high density of
nitride fuels can bring out more excess neutrons, therefore, they
have a higher potential to transmute the long lived fission
products. As for reprocessing feasibility, actinide nitrides also
appear to be a compromise between oxide and metal fuels. For the
sake of better understanding of the behavior of these materials
under irradiation, their accurate electronic structure description
by first-principles methods is necessary. Actinides form an
isostructural series of mononitrides with a simple rock-salt type
structure and a complete solid solubility in the whole composition
range. In virtue of the prospective use of actinide mononitrides as
advanced fuel materials, it is of crucial importance to know their
thermal properties for modeling the fuel behavior at elevated
temperatures. The thermodynamic properties such as standard
enthalpies of formation and heat capacities are essential to predict
the phase stability including the melting points.

Despite the abundant research on actinide mononitride, however, for ThN
compound, relatively little is known regarding its chemical bonding,
mechanical properties, high pressure melting points, and phonon dispersion. In
the early theoretical studies, only the lattice parameters and bulk modulus of
the actinide nitride series have been calculated by utilizing linear
muffin-tin orbital (LMTO) method \cite{Brooks}. Although the electronic
properties and chemical bonding in ThN have been recently calculated by Shein
\emph{et al}. \cite{Shein} through the full-potential
linear-augmented-plane-wave (FLAPW) method, the study of the bonding nature of
Th-N ionic/covalent character is still lacking. These facts, as a consequence,
inhibit the fundamental understanding of thorium momonitride from basic point
of view. Motivated by these observations, in this paper, we present a
first-principles density functional theory (DFT) study by calculating the
structural, electronic, mechanical, and thermodynamic properties of ThN. As
will be shown, a preliminary study of some bulk properties of ThN indicates
that the conventional generalized gradient approximation (GGA) for the
exchange-correlation potential in DFT can give satisfactory results when
compared to experimental data and thus does not require other treatments
beyond LDA/GGA such as LDA/GGA+\emph{U}, which are indispensable in the
uranium and plutonium compounds for stronger localization and correlation of
their 5\emph{f} electrons.

After optimizing the ground state structure of ThN, we performed the
calculation of density of states (DOS) and Bader analysis \cite{Bader} for
ThN. The results indicate that the Th 5\emph{f} states are involved in the
formation of Th-N and Th-Th interatomic bonds and about 1.82 electrons
transfer from each Th atom to its surrounding N atoms. Through the mechanical
analysis, we find that NaCl-style ThN is mechanically stable and elastically
anisotropic. In order to predict the melting curve, we calculated the elastic
constants, bulk modulus and shear modulus from ambient pressure to 117 GPa.
Since we have known that ThN melts congruently at 2790$\pm$30$^{\circ}$ under
a nitrogen pressure somewhat less than 1 atm \cite{Olson}, thus the ThN can be
ranked as refractory materials. Utilizing the Lindemann melting criterion
\cite{Lindemann}, we got the melting curve versus pressures. The melting
points were enhanced by about 2100 K from ambient pressure to 117 GPa. Our
predicted results indicate that ThN is able to withstand temperatures above
5100 K without chemical change and physical destruction under high pressures.
The calculated phonon dispersion confirms the dynamic stability for ThN. Based
on our phonon dispersion data, the lattice vibration energy, thermal
expansion, and specific heat are obtained by using the quasiharmonic
approximation (QHA). Our calculated special heat, including both lattice and
conduction electron contributions, agrees well with experimental results at
\emph{T}$<$1500 K domain.

The rest of this paper is organized as follows. The first-principles
computational details are briefly introduced in Sec. II. The
calculated results are presented and discussed in Sec. III. Finally,
a summary of this work is given in Sec. IV.

\section{computational methods}

The first-principles total energy calculations were carried out using the
Vienna \textit{ab initio} simulations package (VASP) \cite{G.Kresse2} with the
projected-augmented-wave (PAW) pseudopotentials \cite{PAW} and plane waves.
The exchange and correlation effects were described within GGA \cite{GGA}. The
thorium 6\emph{s}$^{2}$7\emph{s}$^{2}$6\emph{p}$^{6}$6\emph{d}$^{1}$%
5\emph{f}$^{1}$ and nitrogen 2\emph{s}$^{2}$2\emph{p}$^{3}$ electrons were
treated as valence electrons. The electron wave function was expanded in plane
waves up to a cutoff energy of 500 eV. We have performed numerous convergence
tests on determining the influence of the \emph{k}-point mesh on the total
energy. The Monkhorst-Pack \cite{Monkhorst} 11$\times$11$\times$11 mesh was
used in Brillouin zone (BZ) integration, which turns to be sufficient to get
results converged to less than 1.0$\times$10$^{-4}$ eV per atom. The
corresponding electronic density of states (DOS) was obtained with 19$\times
$19$\times$19 \emph{k}-point mesh.

It is known that the elastic constants are defined by means of a Taylor
expansion of the total energy, E(\emph{V},$\delta$), for the system with
respect to a small strain $\delta$ on the equilibrium cell according to the
following law \cite{Fast}:
\begin{equation}
E(V,\delta)=E(V_{0},0)+V_{0}\left[  \sum_{i}\tau_{i}\xi_{i}\delta_{i}+\frac
{1}{2}\sum_{ij}C_{ij}\delta_{i}\xi_{j}\delta_{j}\right]  , \label{delta}%
\end{equation}
where \emph{E}(\emph{V$_{0}$}, 0) and \emph{V$_{0}$} are the total energy and
volume of the equilibrium cell without strains, respectively, $\tau_{i}$ is an
element in the stress tensor, and $\xi_{i}$ is a factor presenting to take
care of the Voigt index \cite{Voigt}.

\begin{table}[ptb]
\caption{Strains used to calculate the elastic constants of NaCl-type ThN.}%
\begin{ruledtabular}
\begin{tabular}{cccccccccccccccc}
Strains&Parameters (unlisted: $\varepsilon_{ij}$=0)&$\frac{1}{V}\frac{\partial^{2} E(V,\delta)}{\partial \delta^{2}}$$|_{\delta=0}$\\
\hline
1&$\varepsilon$$_{11}$=$\delta$&\emph{C}$_{11}-\emph{P}$\\
2&$\varepsilon$$_{11}$=$\varepsilon$$_{22}$=$\delta$&2(\emph{C}$_{11}$+\emph{C}$_{12}-\emph{P}$)\\
3&$\varepsilon$$_{13}$=$\varepsilon$$_{31}$=$\delta$&4\emph{C}$_{44}-2\emph{P}$\\
\end{tabular}
\label{strain}
\end{ruledtabular}
\end{table}To calculate the elastic constants of rock-salt structure ThN, we
applied three independent strains. The parametrizations that we used for these
strains are given in Table \ref{strain}. We calculated the total energy of
each strain for a number of small values of $\delta$. These energies were then
fitted to a polynomial in $\delta$ and the curvatures of the energy versus
$\delta$ curve were obtained for using in Eq. (1). The elastic constants are
closely related to many physical properties of solids, such as the Debye
temperature, specific heat, melting temperature, and Gr\"{u}neisen parameter,
etc. At low temperatures, the vibrational excitations arise solely from
acoustic vibrations. Hence, the Debye temperature calculated from elastic
constants is the same as that determined from specific heat measurements at
low temperature. The relation between the Debye temperature ($\theta_{D}$) and
the average elastic wave velocity ($v_{m}$) is
\begin{equation}
\theta_{D}=\frac{h}{k_{B}}(\frac{3n}{4\pi\Omega})^{1/3}v_{m}, \label{debye}%
\end{equation}
where \emph{h} and $k_{B}$ are Planck and Boltzmann constants, respectively,
\emph{n} is the number of atoms in the molecule and $\Omega$ is molecular
volume. The average elastic wave velocity in the polycrystalline materials is
approximately given by
\begin{equation}
v_{m}=\left[  \frac{1}{3}(\frac{2}{v_{t}^{3}}+\frac{1}{v_{l}^{3}})\right]
^{-1/3}, \label{vm}%
\end{equation}
where $v_{t}$=$\sqrt{G/\rho}$ ($\rho$ is the density) and $v_{l}$%
=$\sqrt{(3B+4G)/3\rho}$ are the transverse and longitudinal elastic wave
velocities, respectively.

After obtaining the Debye temperature at various pressures, furthermore, we
performed the melting curve calculation from Debye temperatures. The
calculation was based on the Lindemann melting criterion \cite{Lindemann}.
This model is based on the harmonic approximation, predicting that melting
will occur when the ratio of the root mean square (rms) atomic displacement to
the mean interatomic distance reaches a certain value (generally about 1/8).
It can be expressed as follows
\begin{equation}
T_{m}=CV^{2/3}\theta_{D}, \label{Lindemann}%
\end{equation}
where $T_{m}$ is the melting point, $C$ is a constant, and $V$ is atomic volume.

\section{results and discussions}

\subsection{Atomic and electronic structures}

In order to examine the possibility of magnetism in ThN, the spin-polarized
calculations by assuming initial ferromagnetic state were carried out. We
found that the ground state for ThN is non-magnetic, without any localized
atomic magnetic moments, the result of which is in agreement with observed
\cite{Kleykamp} paramagnetism in ThN. In this paper, the theoretical
equilibrium volume \emph{V$_{0}$}, bulk modulus \emph{B} and the pressure
derivative \emph{B}$^{^{\prime}}$ are obtained by fitting the third-order
Brich-Murnaghan equation of state (BMEOS) \cite{Brich}. Our calculated lattice
parameter for the cubic unit cell of ThN is \emph{a}$_{0}$=5.179 {\AA }, which
is in good accordance with the experimental data of 5.16 {\AA } \cite{Wedwood}%
. The bulk modulus \emph{B} and its pressure derivative \emph{B}$^{^{\prime}}$
obtained by fitting the BMEOS are 176.1 GPa and 3.9, respectively, which are
also consistent with the corresponding experimental values of 175 GPa and 4.0
\cite{Gerward}. The calculated \emph{B} for ThN is much larger than that of
metallic $\alpha$-Th (about 60-62 GPa \cite{Hachiy}), i.e., a pronounced
increase of structural rigidity from metal to nitride due to the direct Th-N
bonding formation, which occurs as well as for other NaCl-type metal mononitrides.

The total DOS and site-projected orbit-resolved DOS (PDOS) of ThN are
displayed in Fig. 1. Evidently, the conduction band is strongly marked by Th
5\emph{f} states, with a little bit degrees of Th 6\emph{d} and N 2\emph{p}
states. The valence band ranging from $-$5.5 eV to $-$1 eV is of mixed Th
\emph{d}$/$\emph{f} and N \emph{p} character. Near the Fermi level are
composed by comparable contributions of Th 6\emph{d} and 5\emph{f} states,
whereas the contributions of N 2\emph{p} states become much smaller. Due to
the observable \emph{d} and \emph{f} states near the Fermi energy, ThN
exhibits a clear metallic behavior.

\begin{figure}[ptb]
\begin{center}
\includegraphics[width=0.5\linewidth]{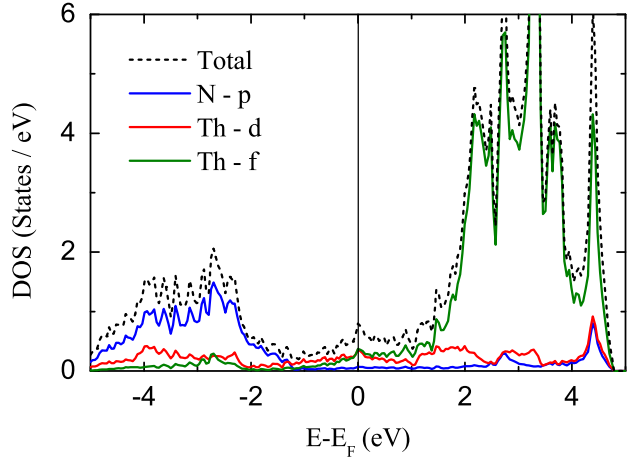}
\end{center}
\caption{(Color online) Total and site-projected orbital-resolved electronic
densities of states for ThN at equilibrium. The Fermi energy is set at zero}%
\end{figure}

\begin{figure}[ptb]
\begin{center}
\includegraphics[width=0.4\linewidth]{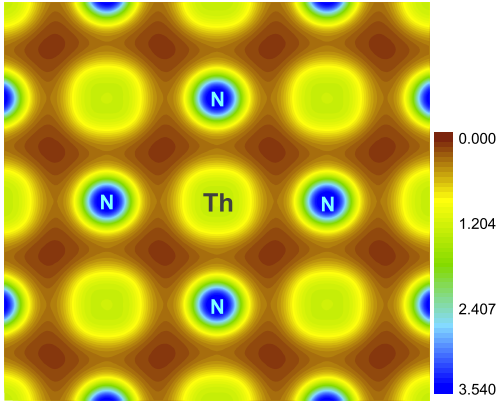}
\end{center}
\caption{(Color online) Valence charge density of ThN in (100) plane.}%
\label{charge}%
\end{figure}

In order to further analyze the chemical bonding nature, we display in Fig.
\ref{charge} the charge distribution in ThN (100) plane and list in Table
\ref{bader} the Bader effective charges. We find the near-spherical
distribution of electron density around thorium and nitrogen atoms, with
rather small density value between them. This is typical for crystals with the
rock-salt structure having the ionic bonding due to the charge transfer from
metal to non-metal atoms. Since Bader analysis \cite{Bader} is an effective
tool for studying the topology of the electron density and suitable for
discussing the ionic/covalent character of a compound, thus we also performed
the Bader effective charge calculation for ThN. For this we adopted
300$\times$300$\times$300 charge density grids, then the spacing between
adjacent grid points is 0.0172 {\r{A}}. The calculated effective atomic
valance charges and volumes are listed in Table \ref{bader} together with the
UN and PuN results for comparison. Our results show that about 1.82 electrons
transfer from each Th atom to its neighboring N atoms. The effective valency
in ThN then can be represented as Th$^{+1.82}$ N$^{-1.82}$, while UN is
U$^{+1.71}$ N$^{-1.71}$ \cite{Lu} and PuN is Pu$^{+1.59}$ N$^{-1.59}$,
indicating that the ionicity in ThN is somewhat stronger than that in UN and
PuN, as shown in Table II.

\begin{table}[ptb]
\caption{Calculated effective atomic charge and volumes according to Bader
partitioning of AnN (An=Th, U, Pu).}%
\begin{ruledtabular}
\begin{tabular}{cccccccccccccccc}
Compound&Q$_{B}$(\emph{A})&Q$_{B}$(N)&V$_{B}$(\emph{A})&V$_{B}$(N)\\
&(\emph{e})&(\emph{e})&(\AA$^{3}$)&(\AA$^{3}$)\\
\hline
ThN&+1.82&-1.82&20.924&13.848\\
UN&+1.71&-1.71&18.301&11.581\\
PuN&+1.59&-1.59&17.713&11.757\\
\end{tabular}
\label{bader}
\end{ruledtabular}
\end{table}

\subsection{Mechanical properties}

\begin{figure}[ptb]
\begin{center}
\includegraphics[width=0.8\linewidth]{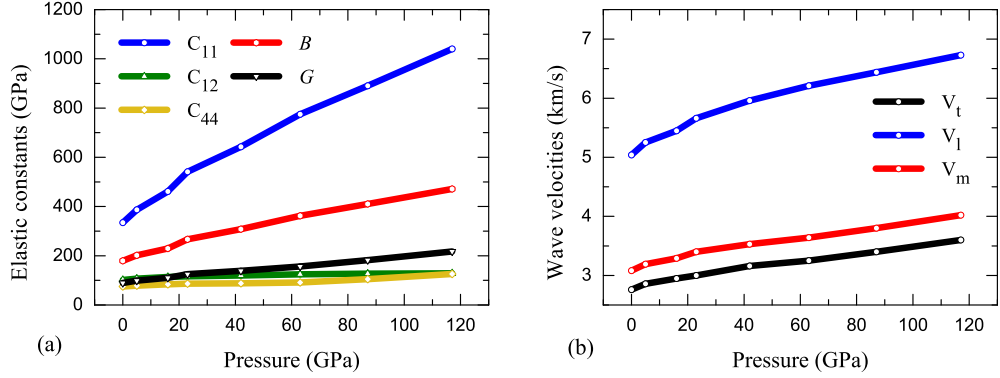}
\end{center}
\caption{(Color online) (a) Evolution with pressure of the ThN elastic
constants, bulk modulus and shear modulus; (b) Evolution with pressure of the
transverse elastic wave velocity $v_{t}$, longitudinal wave velocity $v_{l}$,
and the average wave velocity $v_{m}$.}%
\label{B-P}%
\end{figure}

For the cubic structure, there are three independent elastic constants
$C_{11}$, $C_{12}$, and $C_{44}$. Under pressure \emph{P}, the relation
between the elastic constants \emph{C}$_{ij}$ and the bulk modulus \emph{B}
can be expressed as follows \cite{Sinko}:
\begin{equation}
C_{11}+2C_{12}=3B-P.
\end{equation}
At zero pressure, the well-known expressions of Voigt and Reuss
\cite{Voigt,Reuss,Preston} bulk moduli can be re-derived. Under the
Voigt approximation, the effective shear modulus \emph{G}$_{V}$ for
cubic phase can be expressed as
$G_{V}=\frac{C_{11}-C_{12}+3C_{44}}{5}$, while with Reuss
approximation the shear modulus can be expressed as, $G_{R}=\frac
{5(C_{11}-C_{12})C_{44}}{4C_{44}+3(C_{11}-C_{12})}$. Hill
\cite{Hill} proved that the Voigt and Reuss equations represent
upper and lower limits of the true polycrystalline constants, and
recommended that the shear modulus \emph{G} is an arithmetic average
of Voigt and Reuss approximations, i.e.,
$G=\frac{1}{2}(G_{R}+G_{V})$. From that, the Young's modulus
\emph{E} and
Poisson's ratio $\nu$ can be given by $E$=$\frac{9BG}{3B+G}$ and $\nu$%
=$\frac{3B-2G}{2(3B+G)}$. Using the above functions, the calculated bulk
modulus \emph{B}, the pressure derivative of bulk modulus \emph{B$^{^{\prime}%
}$}, shear modulus \emph{G}, Young's modulus \emph{E}, and Poisson's ratio
$\nu$ of ThN are given in Table \ref{c}. For comparison, the experimental data
from Ref. \cite{Gerward} and the theoretical FLAPW-GGA results in Ref.
\cite{Shein2} are also presented. As can be seen from Table \ref{c}, our
calculated $C_{12}$ and $C_{44}$ are in agreement with the FLAPW-GGA results,
while $C_{11}$=334.8 GPa is smaller than the corresponding data 396.6 GPa. At
ground state, the bulk modulus \emph{B} derived from elastic constants is
179.4 GPa, which is well consistent with that obtained by BMEOS fitting. This
value is very close to the experimental data of 175 GPa \cite{Gerward} but
smaller than the FLAPW-GGA result of 199.9 GPa. The pressure derivative of the
bulk modulus \emph{B$^{^{\prime}}$} is 3.9, which is also in good accordance
with the experimental data of 4.0. The calculated Poisson's ratio $\nu$=0.286
is consistent with the FLAPW-GGA result. With a cubic system, the elastic
anisotropy factor is given by $A$=$\frac{2C_{44}}{C_{11}-C_{12}}$. Since that
micro-cracks in materials can be easily induced by significant elastic
anisotropy, so it is important to evaluate anisotropic factors in
understanding their mechanical durability. Our present value of $A$ is equal
to 0.64, indicating that ThN is elastically anisotropic. As for isotropic
crystals the factor $A$ is equal to 1.0, while with any value larger or
smaller than 1.10 indicating elastic anisotropy.

Besides, Figure \ref{B-P}(a) shows the pressure dependence of the elastic
constants, bulk modulus \emph{B}, and shear modulus \emph{G}. One can clearly
see that the elastic constants all linearly increase with pressure. The value
of $C_{11}$, $C_{12}$ and $C_{44}$ are enhanced by 705 GPa, 27 GPa and 51 GPa,
respectively, from 0 GPa to 117 GPa. These elastic constants satisfy the
generalized elastic stability criteria for cubic crystals under pressure,
i.e., \emph{\~{C}}$_{11}$ $>$ $|$\emph{\~{C}}$_{12}$$|$, \emph{\~{C}}$_{11}$ +
2$|$\emph{\~{C}}$_{12}$$|$ $>$ 0, \emph{\~{C}}$_{44}$ $>$ 0, where
\emph{\~{C}}$_{ii}$=$C_{ii}\mathtt{-}$\emph{P} ($i$=1, 4) and \emph{\~{C}%
}$_{12}$ = $C_{12}$+\emph{P}. The dependence of bulk modulus \emph{B} and
shear modulus \emph{G} on pressure also linearly increase with pressure. These
pressure dependence of elastic constants will be used in calculating the Debye
temperature, and further predicting the melting temperatures at elevated pressures.

\begin{table}[ptb]
\caption{Calculated elastic constants, elastic moduli, pressure derivative of
the bulk modulus \emph{B$^{^{\prime}}$}, Poisson$^{\prime}$s ratio $\upsilon$,
and anisotropic factor \emph{A} for ThN at 0 GPa. Except \emph{B$^{^{\prime}}%
$}, $\upsilon$ and \emph{A}, all other values are in units of GPa.}%
\begin{ruledtabular}
\begin{tabular}{cccccccccccccccc}
Method&\emph{C}$_{11}$&\emph{C}$_{12}$&\emph{C}$_{44}$&\emph{B}&\emph{B$^{ '}$}&\emph{G}$_{V}$&\emph{E}&\emph{$\upsilon$}&\emph{A}\\
\hline
GGA&334.8&101.7&75.0&179.4&3.9&89.5&230.3&0.286&0.64\\
FLAPW-GGA$^{a}$&396.6&101.5&79.9&199.9&&102.4&262.5&0.281&0.54\\
Expt.$^{b}$&&&&175&4.0&\\
\end{tabular}
\label{c}
\end{ruledtabular}
$^{a}$\cite{Shein2}, $^{b}$\cite{Gerward}\end{table}

\subsection{Melting curve of ThN}

\begin{figure}[ptb]
\begin{center}
\includegraphics[width=0.5\linewidth]{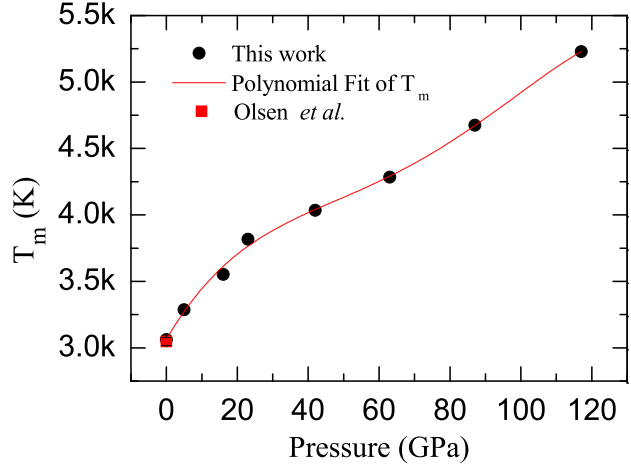}
\end{center}
\caption{(Color online) The melting temperature of ThN as a function of
pressure.}%
\label{tm}%
\end{figure}

The melting curves of materials have great scientific and technological
interest. To understand the transition between solid and liquid phases of ThN
at high pressures, we will use the Lindemann criterion [Eq. (\ref{Lindemann})]
to calculate the melting curve. The Lindemann criterion will be used here as a
single-parameter model and the free constant \emph{C} can be calculated from a
single point. In order to perform the evolution of $T_{m}$, we should know the
melting temperature $T_{m}$ at \emph{P}=0 GPa and the evolution of $\theta
_{D}$. On the one hand, we have known that ThN melts congruently at 2790$\pm
$30$^{\circ}$ under a nitrogen pressure somewhat less than 1 atm \cite{Olson}.
On the other hand, using the above calculated elastic constants, we can derive
the value of transverse and longitudinal elastic wave velocities, i.e.,
$v_{t}$ and $v_{l}$, and then we can get the average wave velocity $v_{m}$ by
Eq. (\ref{vm}). From the relation between $\theta_{D}$ and $v_{m}$ [Eq.
(\ref{debye})], we can finally get the value of $\theta_{D}$. Figure \ref{B-P}
(b) shows the pressure dependence of the elastic wave velocities. One can see
that all the three wave velocities increase with augmenting pressure. Through
the study of evolution of Debye temperature, we can calculate the melting
temperatures using the Lindeman criterion. As is shown in Fig. \ref{tm}, the
melting points of ThN have been calculated from ambient pressure to 117 GPa.
It can be seen that the melting temperature of ThN is increased by about 2100
K from ambient pressure to 117 GPa.

\subsection{Formation energy}

The formation energy of a specific compound is defined as the difference
between the total energy of the compound and of its constitutive elements. The
composition reaction of ThN is as follows,
\begin{equation}
\text{Th}+\frac{1}{2}\text{N}_{2}\rightarrow\text{ThN,}%
\end{equation}
which yields the following expression for the formation energy:
\begin{equation}
E_{f}(\text{ThN})=E(\text{ThN})-(E_{\text{Th}}+\frac{1}{2}E_{\text{N}_{2}%
})\text{.}%
\end{equation}
In order to calculate the formation energy \emph{E$_{f}$}, the total energy of
ThN, $\alpha$-Th and N$_{2}$ dimer should be calculated. Density-functional
theory is known to overestimate the binding energy \emph{E$_{b}$} of N$_{2}$
dimer, so it will result in an underestimation of the present reaction energy
via the \emph{E}$_{\mathtt{N}_{2}}$ term. However, this error can be remedied
by shifting the energy of N$_{2}$ so as to give the experimental binding
energy. The experimental \emph{E$_{b}$} of N$_{2}$ is 9.9 eV \cite{Stampfl}.
In the GGA the \emph{E$_{b}$} is overestimated by about 0.7 eV/N$_{2}$. The
formation energy \emph{E$_{f}$} is $-$2.14 eV/atom with the correction in the
\emph{E}$_{\text{N}_{2}}$ term. This value is somewhat higher than the
experimentally meaured enthalpy of formation $\Delta_{f}H$=$-$1.81 eV/atom at
room temperature \cite{Sed}. Since the formation energy calculated here does
not take into account the temperature, therefore, it can only give insight
into the stability of the compounds at low temperatures.

\subsection{Phonon dispersion curve}

\begin{figure}[ptb]
\begin{center}
\includegraphics[width=0.5\linewidth]{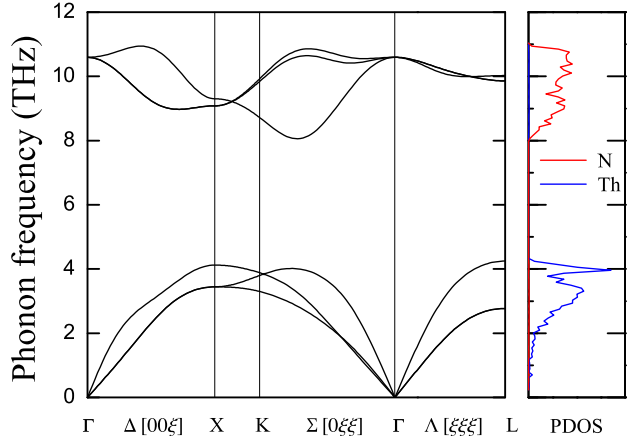}
\end{center}
\caption{(Color online) Calculated phonon dispersion curves (left panel) and
corresponding projected phonon DOS (right panel) for ThN.}%
\label{phonon}%
\end{figure}In calculating the phonon dispersion curves and the phonon density
of states, the Hellmann-Feynman theorem and the direct method \cite{Parlinski}
are employed. For the BZ integration, the 3$\times$3$\times$3 Monkhorst-Pack
\emph{k}-point mesh is used for the 2$\times$2$\times$2 ThN supercell which
contains 64 atoms. In order to calculate the Hellmann-Feynman forces, we
displace two atoms (one Th and one N atoms) from their equilibrium positions
and the amplitude of all the displacements is 0.03 \AA . The calculated phonon
dispersion curves along $\Gamma-X-K-\Gamma-L$ directions is displayed in Fig.
\ref{phonon}. For rock-salt type ThN, there are only two atoms in its formula
unit, therefore, six phonon modes exist in the dispersion relations. The
projected phonon DOS for ThN is also plotted in Fig. \ref{phonon}. Because of
the fact that the thorium atom is heavier than the nitrogen atom, the phonon
DOS splits into two parts with an evident gap: one part is in the range of 0-4
THz where the vibrations of thorium atoms are dominant; the other part is in
the domain of 8-11 THz where the vibrations mainly come from nitrogen atoms.
Our calculated optical frequency at $\Gamma$ is about 10.6 THz, and this value
is very close to the experimental value of 10.3 THz \cite{Wedwood}. The phonon
dispersion illustrates the stability of ThN, and further indicates that our
following thermodynamic calculations are reliable.

\subsection{Thermodynamic properties}

\begin{figure}[ptb]
\begin{center}
\includegraphics[width=0.5\linewidth]{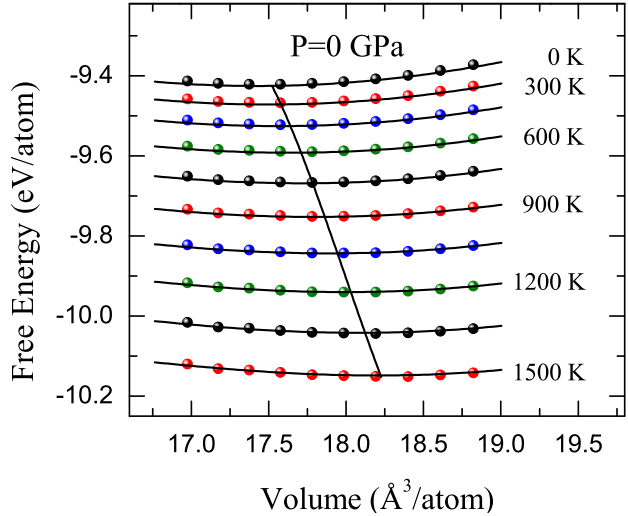}
\end{center}
\caption{(Color online) Dependence of the Helmholtz free energy \emph{F}%
(\emph{T,V}) on crystal volume at various temperatures and the locus of the
minimum of the free energy for ThN.}%
\label{helmholtz}%
\end{figure}\begin{figure}[ptbptb]
\begin{center}
\includegraphics[width=0.5\linewidth]{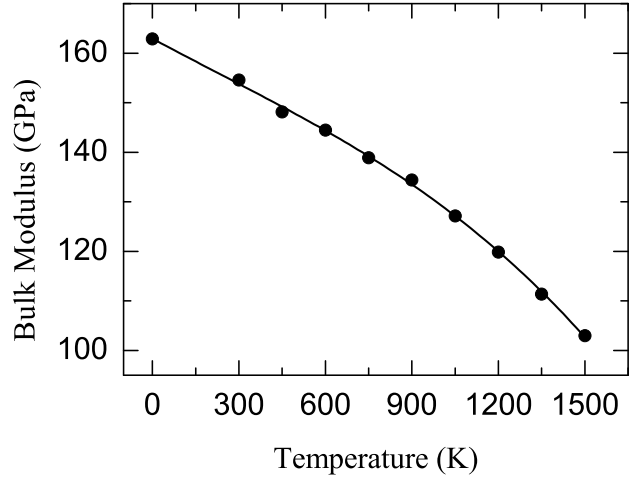}
\end{center}
\caption{(Color online) Temperature dependence of the bulk modulus for ThN.}%
\label{bulk}%
\end{figure}

\begin{figure}[ptb]
\begin{center}
\includegraphics[width=0.6\linewidth]{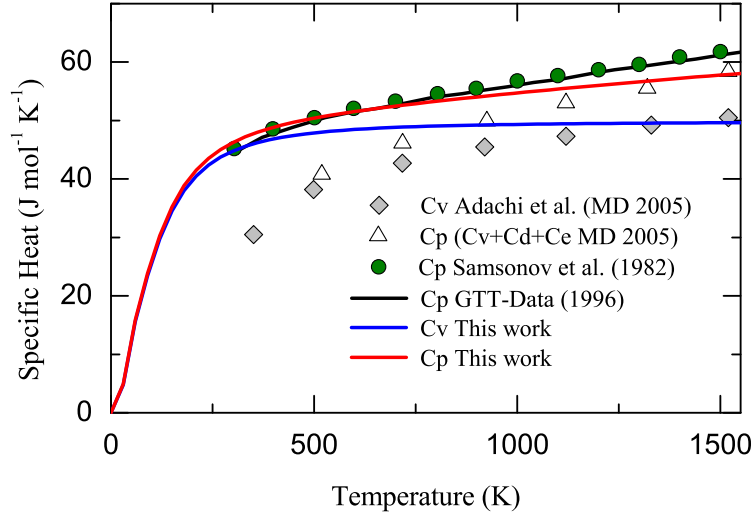}
\end{center}
\caption{(Color online) Specific heat capacities of ThN. Experimental data
from Refs. \cite{1982,1996} and theoretical results from Ref. \cite{Adachi}
are also displayed for comparison.}%
\label{cp}%
\end{figure}The Helmholtz free energy \emph{F} in QHA is investigated as
follows:
\begin{equation}
F(V,T)=E(V)+F_{ph}(V,T)+F_{ele}(V,T),
\end{equation}
where $E(V)$ stands for the ground-state cold energy, \emph{F$_{ph}$%
}(\emph{V,T}) is the phonon free energy at a given unit cell volume \emph{V},
and \emph{F$_{ele}$} is electron excitation energy. Under QHA, the
\emph{F$_{ph}$}(\emph{V,T}) can be calculated from phonon DOS $g(\omega)$ by
\begin{equation}
F_{ph}(V,T)=k_{B}T\int_{0}^{\infty}g(\omega)\ln\left[  2\sinh\left(
\frac{\hbar\omega}{2k_{B}T}\right)  \right]  d\omega, \label{pho}%
\end{equation}
where $\omega$=$\omega(V)$ depends on volume and thus Equation
(\ref{pho}) contains some effect of anharmonics. $F_{ele}$ in Eq.
(8) can be obtained from the energy and entropy contributions, i.e.,
$E_{ele}-TS_{ele}$. The electronic entropy $S_{ele}$ is of the form
\begin{equation}
S_{ele}(V,T)=-k_{B}\int{n(\varepsilon,V)[f\ln{f}+(1-f)\ln{(1-f)}]d\varepsilon
},
\end{equation}
where \emph{n}($\varepsilon$) is electronic DOS, and ${{f}}$ is the
Fermi-Dirac distribution. While the energy \emph{E$_{ele}$} due to the
electron excitations takes the following form
\begin{equation}
E_{ele}(V,T)=\int{n(\varepsilon,V)f\varepsilon d\varepsilon}-\int
^{\varepsilon_{F}}{n(\varepsilon,V)\varepsilon d\varepsilon},
\end{equation}
where $\varepsilon_{F}$ is the Fermi energy.

The calculated free energy $F(V,T)$ curves of ThN for temperature ranging from
0 up to 1500 K are shown in Fig. 6. The locus of the equilibrium lattice
parameters at different temperature \emph{T} are also presented. The
equilibrium volume \emph{V(T)} and the bulk modulus \emph{B(T)} are obtained
by BMEOS fitting. Figure 7 shows the temperature dependence of the bulk
modulus \emph{B}. Clearly, the bulk modulus \emph{B(T)} decreases along with
the increase of temperature. This kind of temperature effect is very common
for compounds and metals. Besides, the specific heat at constant volume
$C_{V}$ can be directly calculated through
\begin{equation}
C_{V}=(\frac{\partial F}{\partial T})_{V}=k_{B}\int_{0}^{\infty}%
g(\omega)\left(  \frac{\hslash\omega}{k_{B}T}\right)  ^{2}\frac{\exp
\frac{\hslash\omega}{k_{B}T}}{(\exp\frac{\hslash\omega}{k_{B}T}-1)^{2}}%
d\omega,
\end{equation}
while the specific heat at constant pressure $C_{P}$ can be
evaluated by the thermodynamic relationship
$C_{P}\mathtt{-}C_{V}=\alpha_{V}^{2}(T)B(T)V(T)T$, where the
isobaric thermal expansion coefficient can be calculated according
to the formula $\alpha_{V}(T)$=$\frac{1}{V}\left(  \frac{\partial
V}{\partial T}\right)  _{P}$. Calculated \emph{C}$_{V}$ and
\emph{C}$_{P}$ of ThN are displayed in Fig. \ref{cp}. For
comparison, the experimental data from Refs. \cite{1982,1996} and
the theoretical molecular dynamics (MD) results by Adachi \emph{et
al.} \cite{Adachi} are also presented. In general, our calculated
values of \emph{C}$_{V}$ and \emph{C}$_{P}$ are both higher than the
corresponding MD results up to 1500 K. However, our calculated
\emph{C}$_{P}$, including both lattice and conduction electron
contributions, agrees well with experimental results in a wide
temperature domain with a tiny discrepancy of 4 J/mol$\cdot$K at
1500 K.

\section{conclusion}

In summary, we have performed systematic first-principles
calculations on the structural, electronic, mechanical, and
thermodynamic properties of ThN. Within the GGA method, the ground
state structure of ThN can be well produced. Our calculated lattice
constant is in good accordance with the experimental data, within
0.05\% error. Calculated electronic density of states show the
important contributions of Th 6\emph{d }and 5\emph{f} states to the
Fermi-level occupation. The Bader effective charges of ThN can be
expressed as Th$^{+1.82}$N$^{-1.82}$, which is indicated to be more
ionic when compared to UN and PuN. The mechanical analysis has been
carried out, showing that the rock-salt type ThN is mechanically
stable in a wide range of pressures. Also, we have presented the
melting curve of ThN from ambient pressure to 117 GPa by utilizing
the Lindemann criterion. The calculated phonon dispersion of ThN is
stable, confirming the dynamic stability. Under the QHA method, our
calculated specific heat, including both lattice and conduction
electron contributions, agrees well with experimental results. We
expect that these calculated results will be useful for the
application of thorium nitrides in the Generation-IV reactor and
nuclear industry.

\begin{acknowledgments}
This work was supported by NSFC under Grants No. 51071032 and the
Foundations for Development of Science and Technology of China
Academy of Engineering Physics under Grant No. 2009B0301037.
\end{acknowledgments}

\end{document}